# Response to *Comment on 'Non-representative Quantum Mechanical Weak Values'* by Ben-Israel and Vaidman

B. E. Y. Svensson

*Theoretical High Energy Physics, Department of Astronomy and Theoretical Physics, Lund University, Sölvegatan 14, SE-22362 Lund, Sweden*

E-mail: Bengt_E_Y.Svensson@thep.lu.se

**Abstract**

Ben-Israel and Vaidman (Found Phys (2017). doi:10.1007/s10701-017-0071-x) have raised objections to my arguments that there are cases where a quantum mechanical weak value can be said not to represent the system to which it pertains. They are correct in pointing out that some of my conclusions were too general. However, for weak values of projection operators my conclusions still stand.

## 1  Three main objections

Ben-Israel and Vaidman's [1] raise three main objections to my treatment in [2].

### 1.1 Discontinuous limit

Firstly, Ben-Israel and Vaidman argue that there could be no discontinuity since, in the expression $|in> \otimes |m> - i g S |in> \otimes P_M |m> + O(g^2)$ for the joint system-meter state after a weak measurement, the limit $g \to 0$ is trivially continuous. However, as I have explained at length in [2], in the procedure to infer the weak value, a discontinuity may under certain circumstances occur: in deducing the weak value, one crucial step is to divide by the coupling strength $g$ before taking the limit $g \to 0$. It is therefore the state $S |in>$ *without a factor g* that is relevant to the weak value in the sense that it is its projection onto the postselected state that appears in the weak value. The independence of $g$ implies that this state occurs also for $g = 0$, *i.e.,* for the undisturbed system solely described by the state $|in>$. Thus, if that state $S |in>$ is orthogonal to the state $|in>$ – a situation in [2] called "derailment" by the weak measurement – it is "non-representative" in the sense that it has no overlap with the undisturbed system and therefore does not describe that system.

What is at play here can be seen as the difference between a limit and the value at the endpoint[1], only that now the discontinuity is not a simple functional discontinuity but rather a topological or geometrical one; I give an explicit example at the end of this note.

### 1.2 Non-representative criterion is too general

Secondly, Ben-Israel and Vaidman use the additivity of weak values to rightly point out that the simple criterion $< in |S| in > = 0$ for "non-representativeness" that I proposed cannot be correct in general. It is an intriguing situation: if you apply the standard measurement prescription to arrive at the weak value of $\sigma_z$ *directly*, you have a "derailment". However, if you measure $(\sigma_z)_w$ so to speak "indirectly", by combining measurements of $(\sigma_z \pm \sigma_x)_w$, there is no "derailment". Put otherwise, a weak value – supposed to characterize the system independently of how the weak value is measured – could in fact show properties that depend non-trivially on the very way it is measured. I have investigated this ambiguity from a different angle in [3].

Anyhow, I agree with Ben-Israel and Vaidman that my simple criterion $< in |S| in > = 0$ for "non-representativeness" of *any* weak value is not generally valid. In particular, my critical arguments in [2] against the "Quantum Cheshire cat" example [4] are not valid. One has to be more careful in specifying conditions under which "non-representativeness" occurs.

### 1.3 Projector weak values could be non-representative

To this end, let me focus on weak values of *projectors*. As was pointed out by Vaidman [5], a ("representative") projector weak value has a direct physical meaning in the sense that the presence (or not) of a system in a particular "channel" is measured by a non-vanishing (vanishing) weak value of the projector onto that "channel" (see, *e.g.*, [3] for a detailed description of this "which way" criterion). Such a direct physical meaning entails that the Ben-Israel-Vaidman additivity ambiguity does not occur: add or subtract an operator to a projector and at least one of the ensuing expressions will not be a projector, thus representing a different physical situation.

Furthermore, in order to find examples of non-representativeness, one has to consider more complex situations, allowing for what in [2] is called "subsequent system evolution". This is what Ben-Israel and Vaidman do in their third main remark, where they take issue with me regarding the example of Vaidman's nested Mach-Zehnder interferometer (MZI) setup [5]; see also [2] for details, in particular for the notation. Specifically, they contest my argument for the non-representativeness of a weak value like $(\Pi_B)_w$ of the projector onto one of the arms of the interior MZI.

In this case, I disagree and stick to my conclusions in [2]: the weak value $(\Pi_B)_w$ is "non-representative". As described there, the crucial point is that a weak measurement of $\Pi_B$ does induce a non-zero amplitude for the *E*-arm. This amplitude is proportional to the weak measurement strength *g*. The fact that it is small for small values of *g* is of no importance: the weak value *($\Pi_B)_w$* is obtained from this amplitude *after division by g* (and after postselection). This non-vanishing of the *E*-amplitude is a true example of a "derailment" making the weak value *($\Pi_B)_w$* "non-representative": the undisturbed system has vanishing *E*-amplitude (the *E*-arm is indeed the "dark port" arm for the interior MZI constituted by the *B*- and the *C*-arms). Very explicitly then, the discontinuity in this case is that there is an *E*-amplitude for any $g \neq 0$, but not for *g=0*.

---

[1] An analogy to this situation, although not a perfect one, could be an ordinary function $f(x)$ which for small $x > 0$ has an expansion $f(x) = a x + O(x^2)$. It is continuous at $x = 0$. But now consider instead $(1/x) f(x)$. Then $\lim_{x \to 0} [(1/x) f(x)] = a$, while for $x = 0$ the expression $(1/x) f(x)$ is undefined $= 0/0$.

## 2. Conclusions

I agree with Ben-Israel and Vaidman that the simple criterion $< in\ |S|\ in > = 0$ for "non-representativeness" of *any* observable *S* cannot be upheld. However, restricting to weak values of *projectors*, there are cases – the nested MZI is one – where the "derailment" due to weak measurement results in non-representative weak values.

**Acknowledgement** I am grateful to Johan Bijnens for a careful reading of the manuscript.

.